\documentclass[conference]{IEEEtran}
\IEEEoverridecommandlockouts
\usepackage{cite}
\usepackage{amsmath,amssymb,amsfonts}
\usepackage{algorithmic}
\usepackage{graphicx}
\usepackage{textcomp}
\usepackage{xcolor}
\def\BibTeX{{\rm B\kern-.05em{\sc i\kern-.025em b}\kern-.08em
    T\kern-.1667em\lower.7ex\hbox{E}\kern-.125emX}}
\begin{document}

\title{Central Tendency Bias in Human Selection of AI-Generated Design Variations}

\author{\IEEEauthorblockN{1\textsuperscript{st} Huiyang Chen}
\IEEEauthorblockA{\textit{University of Michigan} \\
Ann Arbor, MI, USA \\
huiyangc@umich.edu}
\and
\IEEEauthorblockN{2\textsuperscript{nd} Keqing Jiao}
\IEEEauthorblockA{\textit{Carnegie Mellon University} \\
Pittsburgh, PA, USA \\
kjiao@andrew.cmu.edu}
}

\maketitle

\begin{abstract}

Image-generation AI systems increasingly support creative work by producing multiple design variations for users to evaluate and select. In such human–AI co-creation workflows, selection becomes a critical stage where human judgment guides AI-generated possibilities toward final outcomes. While presenting multiple alternatives is intended to encourage exploration, the simultaneous multi-option presentation may introduce systematic biases in human decision making.
Drawing on ensemble perception theory, we investigate whether these interfaces induce central tendency bias—the tendency to favor options closer to the center of a design set. We conducted a controlled experiment manipulating the variance of design sets (high vs. low) and measured participants’ selections in both aesthetic preference and representativeness tasks. Results show that higher variance increases the selection of center-proximal designs across both tasks. These findings suggest that multi-variation interfaces in image-generation AI systems may constrain selection diversity, revealing a potential tension between diversity in generated outputs and diversity in human selection outcomes.

\end{abstract}

\begin{IEEEkeywords}
Image-generation AI, Human–AI collaboration, Ensemble perception, Design selection, Central tendency bias
\end{IEEEkeywords}

\section{Introduction}
AI image-generation tools are increasingly embedded in creative workflows as collaborative partners \cite{fang2025, wang2025}. Contemporary tools such as Midjourney, DALL·E and Stable Diffusion often present users with multiple generated outputs that can be compared and selected. This interaction structure positions selection as a critical human-in-the-loop stage that determines which ideas progress and which are discarded.

Despite the growing emphasis on co-creation and collaborative intelligence, most research has focused on improving generation capabilities, enhancing ideation, and integrating AI into design workflows\cite{yang2018,oh2018,fang2025, wang2025,cai2025}. Comparatively less attention has been given to understanding the cognitive processes that shape how humans evaluate and select among generated alternatives. In many generative AI interfaces, outputs are displayed simultaneously in grid-based layouts containing multiple options. While this structure is intended to facilitate comparison and exploration, it may also impact how users internally represent and evaluate alternatives. We draw on ensemble perception theory—the finding that observers can automatically extract summary statistics from groups of simultaneously viewed objects, which can bias judgments toward the set’s central tendency\cite{haberman2009,whitney2018,alvarez2011, corbett2012,habermant2009}.

We propose that generative AI selection contexts may inadvertently trigger similar ensemble mechanisms. When users evaluate multiple AI-generated designs simultaneously, they may form an implicit ``mean representation'' of the set\cite{haberman2009,whitney2018}. If this summary representation influences evaluation, human selection in AI workflows may systematically gravitate toward center-proximal options, even when users aim to choose based on personal preference, innovation potential, or creativity. If higher generation diversity strengthens ensemble effects and reduces selection diversity, generative AI systems may paradoxically produce conservative outcomes despite offering diverse alternatives. 

Understanding this structural interaction between interface design and human cognition is important for building user experiences that expand creative exploration. This study extends prior ensemble perception research to generative AI design contexts by investigating how variance within sets of AI-generated design variations influences human selection.

\section{Related Work}

\subsection{Ensemble Perception and Central Tendency Bias}

Ensemble perception refers to the visual system's ability to rapidly extract summary statistics from groups of objects without explicitly processing each element when multiple objects are viewed concurrently \cite{whitney2018,alvarez2011,corbett2012}. An important factor influencing ensemble perception is the variability within the stimulus set. Higher variance can produce more salient central tendency representations, as greater spread amplifies the contrast between peripheral items and the set center \cite{haberman2009,whitney2018}.

Early research on ensemble perception primarily examined low-level visual features using simple stimuli such as arrays of circles varying in size, bars differing in orientation, or spatial distributions of dots\cite{whitney2018}. Later work demonstrated that ensemble coding can also extend to higher-level stimuli including facial expressions, body postures, and object categories\cite{haberman2009,habermant2009}. However, most prior work has focused on perceptual estimation and memory tasks using relatively simple visual stimuli\cite{brady2011,harrison2021}. Much less is known about whether similar mechanisms operate in evaluative decision-making contexts when users compare and select among multiple complex visual artifacts presented simultaneously in image-generation interfaces.

\subsection{Human-AI Collaboration in Generative Design}
Contemporary generative systems position AI as a variation generator while humans act as evaluators and selectors within creative workflows\cite{yang2018,fang2025,wang2025,cai2025}. Designers typically provide prompts or sketches, after which the system produces multiple candidate outputs that users compare, select, and refine iteratively\cite{yang2018,fang2025,frich2019}. In these workflows, the human role often centers on evaluating sets of generated alternatives and deciding which options progress within the design process.

Recent empirical work has begun to reveal how AI-generated outputs can constrain rather than expand creative exploration. Exposure to AI-generated images during ideation led to higher design fixation and reduced divergent thinking \cite{wadinambiarachchi2024}, suggesting that generative tools may inadvertently narrow the creative space at the idea generation stage.

While such work examines how AI outputs shape the generation of ideas, less attention has been given to how the presentation structure of multiple AI outputs influences human selection behavior. Research in human-computer interaction and decision science shows that the structure of the option presentation, including framing, ordering, and cognitive load, can influence choice behavior  \cite{cockburn2020},\cite{jameson2014}. Yet it remains unclear how the simultaneous multi-option presentation typical of image-generation interfaces interacts with perceptual mechanisms such as ensemble perception. Understanding this interaction is important because it may influence which generated designs users ultimately select and therefore affect the outcomes of human–AI creative collaboration.

\section{User Study}

\subsection{Research Question}
This study investigates how multi-option presentation in generative AI systems may influence design selection behavior. Specifically, we examine whether ensemble perception mechanisms shape how users evaluate and choose among multiple design variations. Our study addresses the following research questions:

\textbf{RQ1:} Do users exhibit central tendency bias when selecting among multiple AI-generated design variations?

\textbf{RQ2:} Does higher variance in design sets increase the likelihood that users select center-proximal designs?

\subsection{Method}
We employed a within-subject experimental design manipulating design set variance (high vs. low) across two selection tasks: a preference task and a representativeness task. The stimuli were organized into 8 design themes. For each theme, 2 sets of 8 poster designs were constructed: a high-variance set and a low-variance set. Each theme therefore contributed 16 designs, and the full stimulus pool contained 128 designs. All variations within a set were displayed simultaneously in a grid layout. Although the stimuli were manually constructed to control stylistic variance, the grid-based multi-option presentation approximates the interaction structure commonly used in contemporary image-generation AI interfaces such as Midjourney and DALL·E.

Image-generation models often produce visually rich but unpredictable variations, making it difficult to systematically control stylistic variance when using model outputs. To isolate the effect of variance from uncontrolled generative randomness, we constructed controlled stimulus sets with matched perceptual centroids. 

For each theme, we first created a larger pool of candidate posters and arranged them along a perceptual style continuum ranging from $-3$  (one stylistic extreme) to $+3$ (the opposite extreme), with 0 representing the perceptual centroid. Across themes, the continuum captured a progression from minimal, sparse compositions to designs with denser graphic elements, richer color combinations, and more complex layouts. Designs were ordered based on holistic perceptual comparison of visual characteristics (e.g., layout proportion, color balance, and element composition), rather than computed numerical metrics. Adjacent scores corresponded to small visual differences along these visual attributes, while larger score differences represented more substantial changes in overall visual composition and complexity.

To construct each continuum, we first identified candidate endpoint designs for each theme (minimal at $-3$ and busy at $+3$), then iteratively selected intermediate designs to form a smooth perceptual progression between the endpoints.

To validate the perceptual ordering, three independent raters (none of whom participated in the main experiment) ranked the designs within each theme along the stylistic continuum. Raters' orderings were highly consistent, with disagreements limited to swaps of adjacent positions. We resolved these minor disagreements through discussion to produce a consensus ordering, which was then used as the basis for stimulus selection.

From the ordered continuum, stimuli were selected to construct the experimental design sets. Two visually similar designs were included at the centroid (0 and 0') to reduce potential stimulus-specific bias associated with a single center design. Although the low-variance sets contained repeated near-centroid positions to reflect clustering, the number of center items used in the dependent measure was held constant across conditions (two designs at 0 and 0').

\textbf{Low-variance sets were constructed by clustering visually similar designs near the centroid:}
\[
\{-1, -1, 0, 0, 0', 0, +1, +1\}
\]

\textbf{High-variance sets sampled the full continuum:}
\[
\{-3, -2, -1, 0, 0', +1, +2, +3\}
\]

As a result, low-variance sets included designs that differed only in subtle adjustments around the central style. High-variance sets covered a broader range of visual configurations spanning the continuum. Figure 1 shows example stimulus sets from one of the design themes illustrating the difference between low-variance and high-variance conditions.

\begin{figure}
    \centering
    \includegraphics[width=1\linewidth]{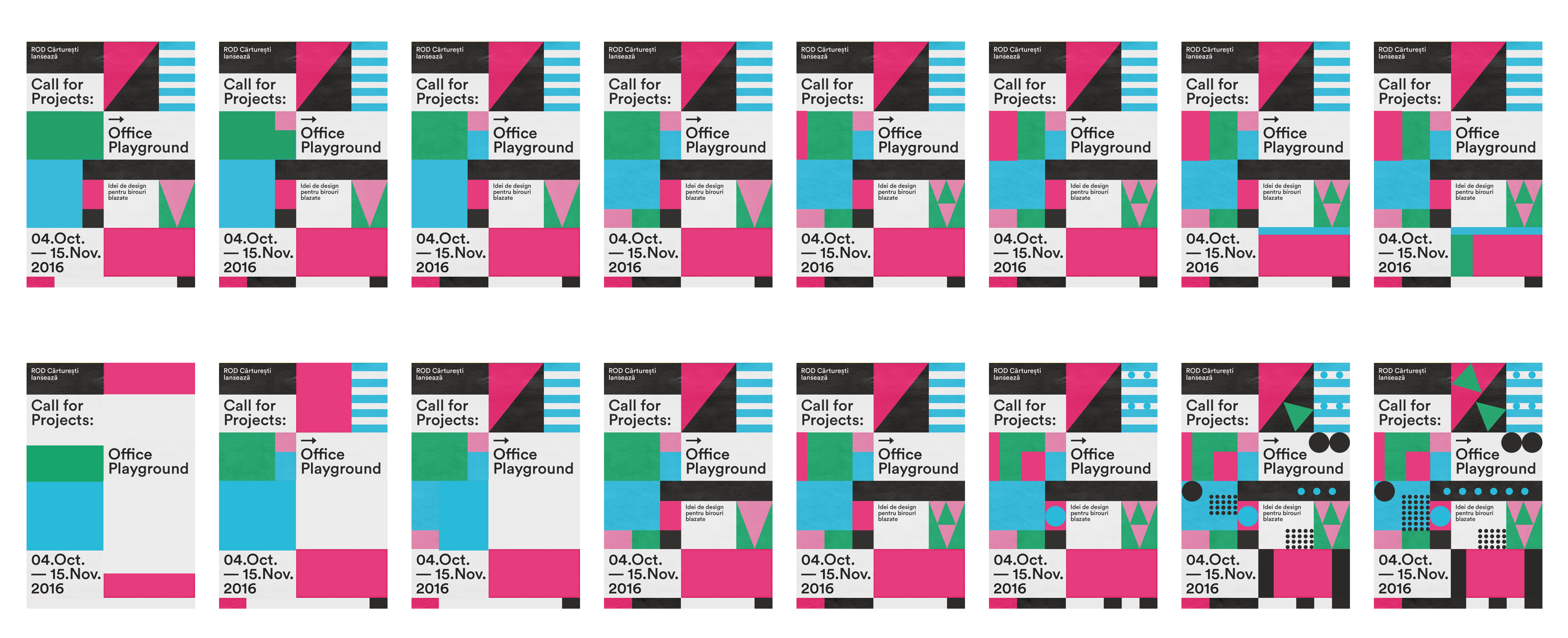}
    \caption{Example stimulus sets used in the experiment. Top row shows a low-variance design set in which designs are visually similar and clustered near the perceptual centroid. Bottom row shows a high-variance design set spanning a broader stylistic continuum. In the actual experiment, designs within each set were presented in randomized grid positions.
    }
    \label{fig:placeholder}
\end{figure}

\subsection{Participants}
Fifty participants (age range 18–35, $M = 23.8$, $SD = 3.4$) were recruited through university mailing lists. All participants reported normal or corrected-to-normal vision and provided informed consent prior to participation. Participants had no prior exposure to the specific design stimuli. Participants had diverse creative backgrounds: 60\% reported some form of design training, 40\% without formal design education. Design experience was recorded but not used as a screening criterion. Exploratory analyses later showed no significant association between design experience and perceived variance ratings.

\subsection{Procedure}
After providing consent, participants received on-screen instructions (5 min) explaining that they would view multiple sets of designs and select one option within each set. To reduce demand characteristics, the instructions avoided mentioning concepts such as ``average,'' ``typical,'' or ``central tendency'', a common practice in studies of ensemble perception and bias. The experiment consisted of two blocks.

Task order was counterbalanced across participants. Half of the participants completed Block 1 (preference task) followed by Block 2 (representativeness task), while the other half experienced the reverse order. Within each block, the order of stimulus sets was randomized for each participant to prevent sequence effects.

No time limit was imposed, but participants were instructed to rely on their intuitive judgment, consistent with prior work on perceptual and aesthetic decision-making.

\textbf{In Block 1}, participants viewed 16 design sets (8 high-variance, 8 low-variance) in randomized order. Each set contained 8 poster designs arranged in a $2\times4$ grid and displayed simultaneously for at least 5 seconds before participants could respond. Within each set, the eight designs were displayed in randomized positions within the $2\times4$  grid to prevent positional bias from confounding ensemble-based central tendency effects. 

Participants then answered the question ``Which design do you like most?'' and selected the primary reason for their choice (more attractive / more creative / more practical / unclear). This phase prioritized ecological validity by positioning preference selection first, before exposure to representativeness framing.

Participants next completed a 1–2 minute filler task consisting of simple arithmetic questions to disrupt potential strategy carry-over.

\textbf{In Block 2}, participants viewed the same 16 sets again in a newly randomized order. For each set they answered the question ``If you could use only one design to represent the overall style of this set, which would you choose?'' This task directly assessed ensemble central-tendency representation by measuring typicality judgments within the design set.

\subsection{Measures}
For each trial, participants selected one design from the eight simultaneously displayed options. Each option corresponded to a position on the global design continuum ranging from $-3$ to $+3$, and selections were coded accordingly. Centralization was quantified as the proportion of trials in which a center design (score 0 or 0’) was selected. We computed each participant's center-selection rate for each variance condition (8 trials per condition), which served as the dependent variable. Finally, participants rated perceived variance on a 1-7 scale (``How different were the designs within each set?''). These ratings served as a manipulation check to confirm the effectiveness of the variance manipulation.

\section{Results}

\subsection{Manipulation Check}
Perceived diversity ratings confirmed successful variance manipulation. Participants rated high-variance sets ($M = 5.71$, $SD = 0.65$) as more diverse than low-variance sets ($M = 3.16$, $SD = 0.72$), $t(49) = 19.57$, $p < 0.001$  (Figure 2). This indicates that participants reliably perceived the intended difference in dispersion between the two conditions. No significant associations were found between perceived diversity ratings and participants' design experience (all $p > 0.10$).

\begin{figure}
    \centering
    \includegraphics[width=1\linewidth]{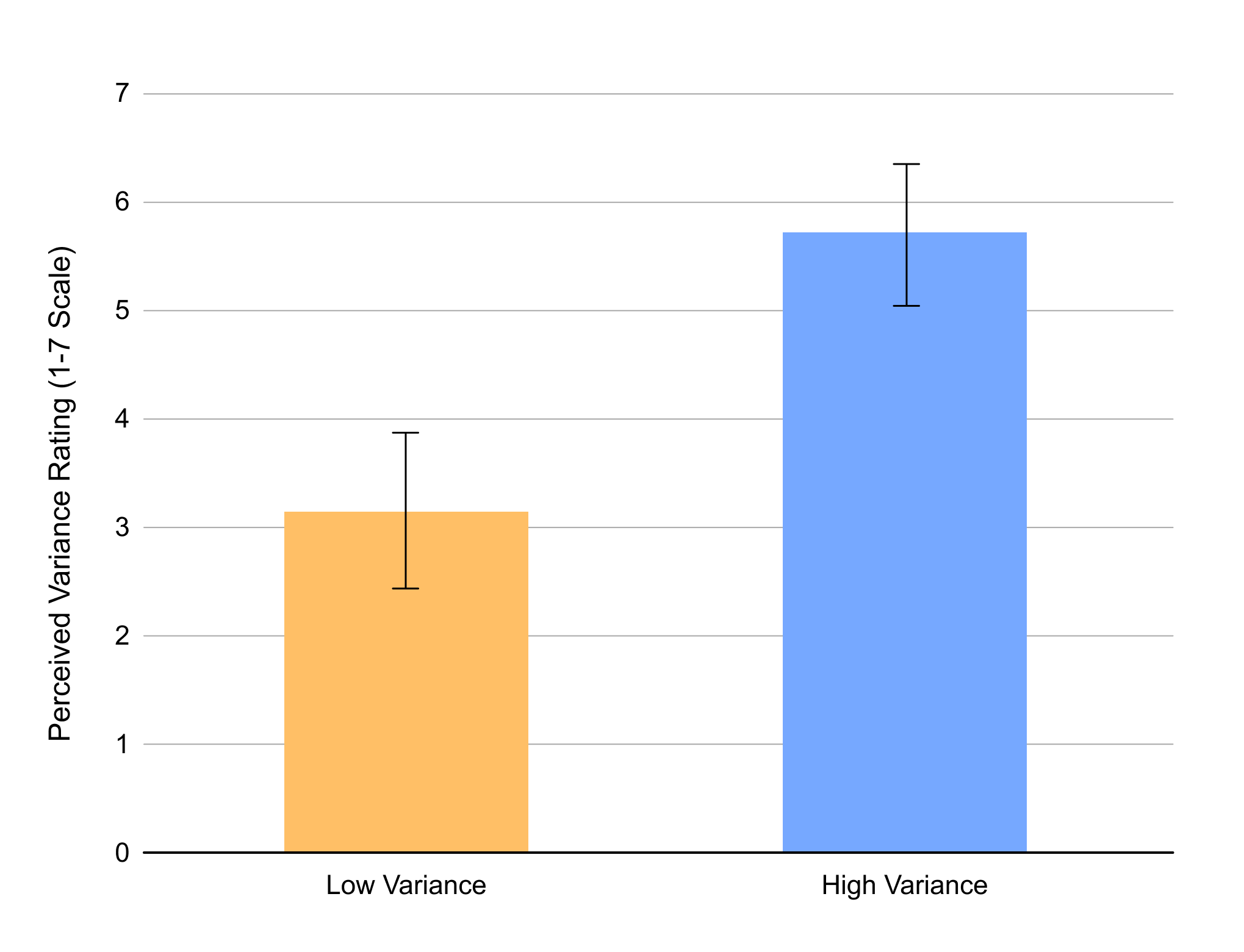}
    \caption{Manipulation Check: perceived variance ratings for high- and low-variance design sets.}
    \label{fig:placeholder}
\end{figure}

\begin{table}[h]
\centering
\caption{Center-selection rates across tasks}
\begin{tabular}{|l|c|c|c|l|}
\hline
Task & High variance & Low variance & $t$(49)&$p$\\
\hline
Preference & 56\%& 34\%& $3.51$&$=0.001$\\\hline
Representativeness & 66\%& 38\%& $4.34$&$<0.001$\\ \hline\end{tabular}

\vspace{1mm}
{\footnotesize
\parbox{0.9\columnwidth}{
Note: All tests are paired-samples t-tests conducted on participant-level center-selection rates (df = 49). Percentages are aggregated across trials (50 participants × 8 sets per condition = 400 trials).
}}
\end{table}

\subsection{Preference task}
In the preference task, participants selected center-proximal designs more frequently in high-variance sets than in low-variance sets. Center designs were chosen on $56\%$ of high-variance trials (224/400) compared to $34\%$ of low-variance trials (136/400). A paired-samples t-test on participant-level center-selection rates revealed a difference between conditions, $t(49) = 3.51, p = 0.001$ (Table I; Figure 3). 

Because 2 of the 8 designs in each set corresponded to center positions (chance = $25\%$), both conditions exceeded chance level, with substantially higher central selection in the high-variance condition.

Participants also reported a primary reason for their preference selection (more attractive / more creative / more practical / unclear). The distribution of reasons was similar for center versus non-center selections (Table II), suggesting that increased center selection under high variance was not accompanied by a clear shift in self-reported selection criteria, consistent with ensemble perception operating as an automatic perceptual mechanism that can influence preference selections independently of conscious evaluative goals.

\begin{table}[h]
\centering
\caption{Self-reported reasons for preference selections (Block 1)}
\begin{tabular}{|l|c|c|}
\hline
Reason& Center selections& Non-center selections\\
\hline
More attractive& 46\%& 43\%\\\hline
More creative& 29\%& 33\%\\\hline

 More practical& 15\%&16\%\\\hline
 Unclear& 10\%&8\%\\ \hline
\end{tabular}

\vspace{1mm}
\begin{flushleft}
{\footnotesize Note: Selections are pooled across both variance conditions (50 participants $\times$ 16 sets = 800 total selections). Center selections include trials in which one of the two shared center designs (position 0 or 0\textquotesingle), present in both high- and low-variance sets, was chosen. Center selections: \textit{n} = 360; non-center selections: \textit{n} = 440. Percentages are rounded to the nearest whole number.}
\end{flushleft}
\end{table}

\begin{figure}
    \centering
    \includegraphics[width=1\linewidth]{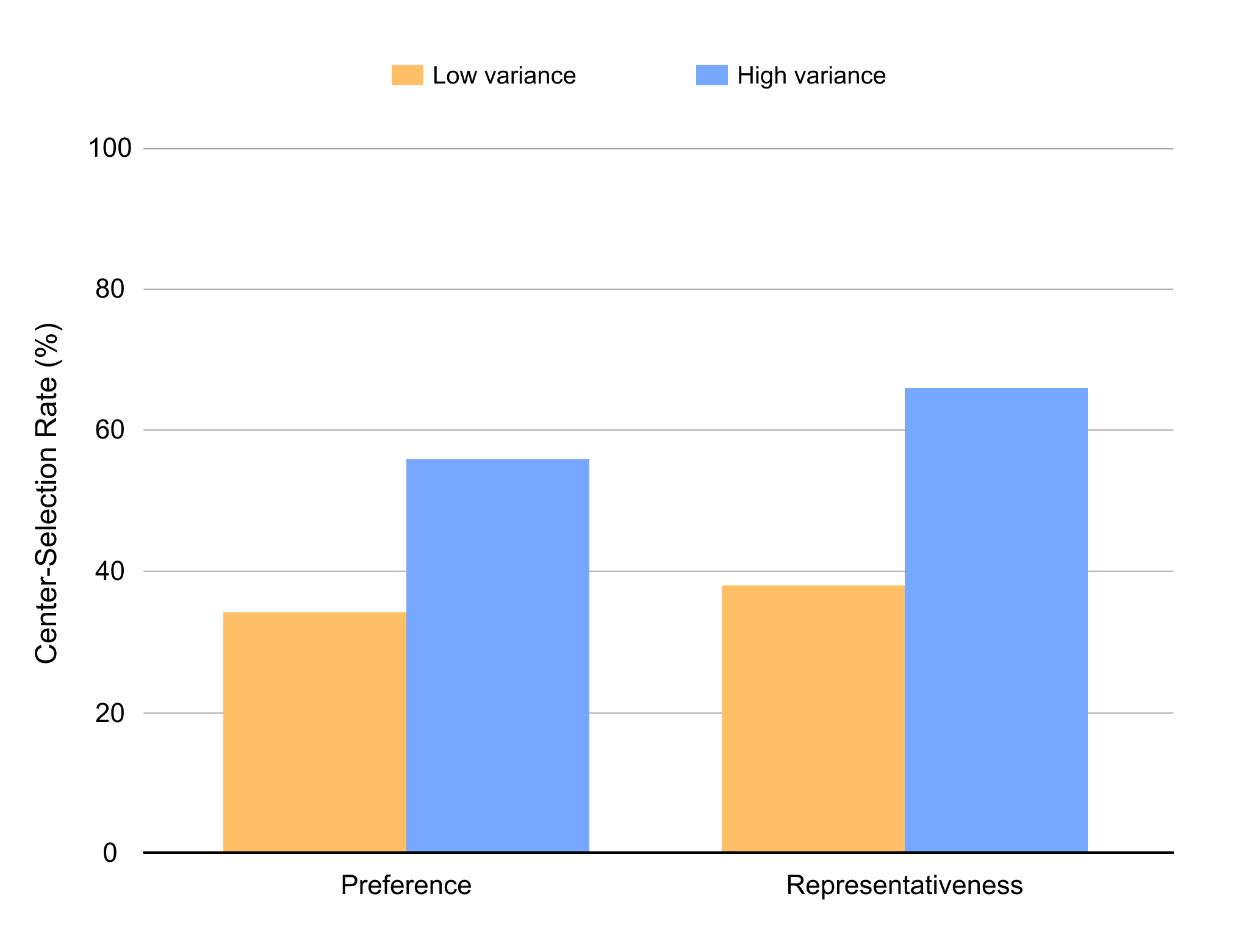}
    \caption{Proportion of center-design selections across variance conditions in preference and representativeness tasks. Percentages are aggregated across trials for descriptive purposes; statistical comparisons were conducted using paired-samples t-tests on participant-level center-selection rates.}
    \label{fig:placeholder}
\end{figure}
\subsection{Representativeness task}
Participants selected center designs on $66\%$ of high-variance trials (264/400) and $38\%$ of low-variance trials (152/400). A paired-samples t-test confirmed a significant difference between conditions, $t(49) = 4.34, p < 0.001$ (Table I; Figure 3).

\section{Discussion and Conclusion}
The results demonstrate that variance within sets of design variations influences user selection behavior, with central tendency bias emerging across both representativeness and aesthetic preference tasks. While centralization in representativeness judgments aligns with ensemble perception theory, its appearance in preference-based decisions is more notable, suggesting that ensemble-based perceptual mechanisms do not merely inform typicality judgments but may also bleed into evaluations users intend to ground in personal taste. When multiple design alternatives are viewed simultaneously, the implicit mean representation of the set appears to function as a cognitive anchor, shaping evaluation even when users are oriented toward subjective liking rather than set-level summarization.

The analysis of self-reported selection reasons provides further evidence for this interpretation. Participants who selected center-proximal designs and those who selected non-center designs reported similar distributions of reasons, with no systematic shift toward any particular evaluative criterion. This suggests that the central tendency bias observed in the preference task did not arise from a conscious change in selection strategy under high variance. Rather, participants appeared to believe they were selecting based on personal aesthetic or creative judgment, while their choices were simultaneously shaped by ensemble-level perceptual mechanisms operating outside conscious awareness. Because users are unlikely to recognize or self-correct for this bias, mitigation strategies must be embedded at the interface level rather than relying on user awareness.

We note that the observed central selection tendency is also consistent with alternative decision-making accounts. Extremeness aversion describes the tendency to avoid options at the extremes of a choice set \cite{chernev2004}, while the compromise effect refers to the increased preference for options that represent intermediate alternatives along key attributes \cite{simonson1992}. While our findings align with predictions from ensemble perception theory, these decision-level frameworks offer complementary explanations that may jointly contribute to the observed behavior. Disentangling the relative contributions of perceptual and decision-level mechanisms remains an important direction for future work.

These findings carry important implications for the design of image-generation AI tools. Many such systems present multiple candidate outputs in grid-based layouts, a structure that may encourage ensemble-based evaluation. Our findings suggest that increasing output diversity may not necessarily yield greater diversity in users' final selection. This points to a potential tension between diversity in generated outputs and diversity in selection outcomes, underscoring the need for interface designs that support broader exploration.

Future generative interfaces could structure the selection stage to counteract ensemble-induced bias. For example, systems might visually emphasize stylistically extreme alternatives within the displayed set by enlarging peripheral options, adding contrast borders, or presenting them in a separate “explore further” panel to draw attention away from the implicit mean and toward novel possibilities. Our results show that central tendency bias persists even under randomized spatial arrangements, suggesting that randomizing spatial positions alone does not eliminate the bias. Interventions may therefore need to go beyond positional changes and consider how diversity is surfaced during selection. One approach is to present subsets of options sequentially rather than simultaneously, which may reduce ensemble extraction over the full set. Such mechanisms could help preserve diversity not only in generated outputs but also in users’ final selections. More broadly, these findings suggest that design evaluations of generative AI tools should explicitly consider selection interface design as a first-class variable alongside generation quality. This shifts the design focus from only how to generate diverse options to how to help users notice, compare, and select across that diversity.

This study has limitations. Because our stimuli were manually constructed rather than generated by commercial image-generation models, the ecological validity with respect to real generative tools may be limited. We made this methodological choice to enable controlled manipulation of set-level variance while holding other visual properties constant, a level of control that would be difficult to achieve with live model outputs. That said, manually constructed stimuli may not fully capture the visual unpredictability and quality variability of real AI-generated outputs. Additionally, the simplified selection tasks focused on aesthetic evaluation and may not reflect more complex design decision contexts. Our sample size (N = 50) may also limit statistical power and the generalizability of findings across broader design domains. Subsequent research should examine ensemble bias in live generative design environments with larger and more diverse participant samples, and explore how presentation structures influence selection behavior across a broader range of creative tasks and workflows. Ultimately, our findings suggest that evaluation of AI creative tools should consider not only generation diversity, but also how outputs are presented during selection, as presentation structures can be consequential for the diversity of final creative outcomes.

\end{document}